\RequirePackage{fix-cm}
\documentclass[twocolumn]{svjour3}          
\smartqed  
\usepackage{graphicx}
\usepackage{amsmath}
\usepackage{enumitem}
\usepackage{graphicx}
\usepackage{dcolumn}
\usepackage{array,multirow}
\usepackage{bm}
\usepackage{amsmath}
\usepackage{epstopdf}
\usepackage{amsfonts}
\usepackage{amssymb}
\usepackage{epsfig}
\usepackage{tabularx}
\usepackage{color}
\usepackage{xcolor}
\usepackage[colorlinks=true,citecolor=blue,urlcolor=cyan,breaklinks]{hyperref}
\usepackage{cite}
\newcommand{\be}{\begin{equation}}
	\newcommand{\ee}{\end{equation}}
\newcommand{\beq}{\begin{eqnarray}}
	\newcommand{\eeq}{\end{eqnarray}}

\begin{document}

\title{The Sagnac effect on a scale--dependent rotating BTZ black hole background
}

\author{\'Angel Rinc\'on \and
        J.R. Villanueva 
}


\institute{\'Angel Rinc\'on \at
	       Instituto de F{\'i}sica, Pontificia Universidad Cat{\'o}lica de Chile, Av. Vicu{\~n}a Mackenna 4860, Santiago, Chile\\
	       \email{arrincon@uc.cl}  
         \and
           J.R. Villanueva \at
           Instituto de F\'isica y Astronom\'ia, Universidad de Valpara\'iso, Avenida Gran Breta\~na 1111, Valpara\'iso, Chile\\
           \email{jose.villanueva@uv.cl}
}

\date{Received: date / Accepted: date}

\maketitle

\begin{abstract}
In this paper we investigate the Sagnac effect by calculating the difference in travel time and phase shift observed for photon beams counter--propa\-gating in  a rotating interferometer on a BTZ black hole solution in the context of scale-dependent gravity, which describes the field around a massive static and rotating object in $2+1$ gravity.
\end{abstract}

\section{Introduction}
\label{intro}
It is very well known that Black Holes (BHs hereafter) are a generic prediction of Einstein's General Relativity (GR). Nowadays, we known that they are more than just simple solutions of Einstein field equations. Of crucial relevance in the black hole theory was the Hawking's work \cite{Hawking:1974rv,Hawking:1974sw} where was shown that black holes indeed emit radiation from their horizon, reason why they are an excellent scenario to study and understand different aspects of gravitational theories. 

Thus, as the Hawking radiation lies at the frontier between GR and Quantum Field Theory (QFT hereafter) \cite{Harmark:2007jy}, a detailed analysis of this and other effects could help us in the search of a complete theory of quantum gravity. 
Despite of Hawking radiation is usually considered as blackbody radiation, it is not completely true. 
The so--called greybody factor measures the modification of the original black body radiation (see \cite{Page:1976df,Unruh:1976fm} for early calculations and \cite{Panotopoulos:2016wuu,Panotopoulos:2017yoe,Panotopoulos:2018pvu,Panotopoulos:2018can,Zhang:2017yfu} for recent applications).

Black holes are described by a few parameters which are: the mass, the angular momentum and the charge. Besides, they have interesting properties which, after the LIGO direct detections of gravitational waves \cite{Abbott:2016blz,Abbott:2016nmj,Abbott:2017vtc}, have received considerable attention. This is the case of the Quasinormal modes (QNM) of black holes, which encoded invaluable information regarding the aforementioned parameters of the solution. For classical review see \cite{Kokkotas:1999bd,Berti:2009kk} and for more recent solutions see \cite{Panotopoulos:2018can,Panotopoulos:2017hns,Destounis:2018utr,Manfredi:2017xcv}.
Given that BHs collect classical and quantum effects, the research of this kind of objects could give us insights about the appropriate way to mix gravity with quantum mechanics.  
%
Gravity in 2+1 dimensions is a perfect background to investigate in detail several well-known effects. The first black hole solution with negative cosmological constant in 2+1 dimensions was the so--called Ba\~nados-Teitelboim-Zanelli (BTZ hereafter) solution which is characterized by mass, angular momentum and charge  \cite{Banados:1992wn,Banados:1992gq}. 

Since its discovery, the BTZ black hole solution has been object of a great amount of investigations based, for example, in its geodesic structure \cite{Farina:1993xw,Cruz:1994ir}, thermodynamics properties \cite{Carlip:1995qv,Banados:1998gg,Cruz:2004vp},  quasinormal modes \cite{Cardoso:2001hn,Crisostomo:2004hj,Panotopoulos:2018can}, stable and regular interior solutions that matches with a BTZ background \cite{gurses,Cruz:1994ar,garcia03,Cruz:2004tz,Garcia:2004jz,cataldo04,Cataldo:2006yk,wuxu}, among others.
In particular, we put our attention in Sagnac effect which is, roughly speaking, ``a comparison of roundtrip times of two light signals traveling in opposite directions along a closed path on a rotating disc" \cite{RAYCHAUDHURI:2014sma}. Our goal here is discuss the Sagnac effect on a scale--dependent rotating BTZ black hole background and compare it with the classical counterpart, as well as analyze if some interesting featuring appears after the inclusion of scale--dependent coupling.  
This article is organized as follow: after this brief introduction, we will discuss the fundamental ingredients of scale--dependent theory of gravity in Sect. \ref{SDT} and, after that, in Sect. \ref{Model} we will discuss briefly the scale--dependent black hole solution in (2+1) dimensions. Then, in Sect. \ref{Sagnac_effect} we perform an analysis of two counter propagating photons to  compute the Sagnac effect on this scale--dependent rotating BTZ black hole solution. Finally, in the Sect. \ref{Conclusions} we will briefly summarize and discuss the main result of this paper.

\section{Scale--Dependent Theory} \label{SDT}
In this section we will summarizes the equations of motion for the
scale--dependent BTZ black hole solution with angular momentum. The idea and notation follows \cite{Koch:2016uso,Rincon:2017ypd,Rincon:2017goj,Rincon:2017ayr,Contreras:2017eza,Rincon:2018sgd,Hernandez-Arboleda:2018qdo,Contreras:2018dhs,Rincon:2018lyd,Rincon:2018dsq,Contreras:2018swc,Contreras:2018gpl,Contreras:2018gct}.
Firstly, the scale--dependent couplings of the theory are two: i) the
Newton’s coupling $G_k$ and ii) the cosmological 
coupling $\Lambda_k$. Notice that the Newton's coupling can be related with the gravitational coupling via $\kappa_k \equiv 8 \pi G_k$. 
What is more, there are two independent fields, they are: i) the metric field $g_{\mu \nu}(x)$ and ii) the scale field
$k(x)$. The effective action is then written as
\begin{align}
\Gamma[g_{\mu \nu}, k] \equiv \int \mathrm{d}^3 x \sqrt{-g}
\Bigg[ 
\frac{1}{2 \kappa_k} \Bigl(R - 2 \Lambda_k \Bigl) \ + \ \mathcal{L}_M
\Bigg]   ,
\end{align}
where $\mathcal{L}_M$ is the Lagrangian density of the matter fields, and the effective Einstein field equations are obtained varying the action respect the inverse metric field, which produce: 
\begin{align}
G_{\mu \nu } + \Lambda_k g_{\mu \nu} \equiv \kappa_k T_{\mu \nu}^{\text{effect}}
\end{align}
where the effective energy momentum tensor is defined according to
\begin{align}
\kappa_k T_{\mu \nu}^{\text{effec}} &=  \kappa_k T_{\mu \nu}^{M} - \Delta t_{\mu \nu}.
\end{align}
Thus, we can understand this effective term as follow: first, the left hand term is the usual contribution given by a matter source whereas the right hand term encoded the scale--dependent effect which incorporate any quantum correction. This new tensor is then defined as:
\begin{align}
\Delta t_{\mu \nu} \equiv G_k \Bigl( g_{\mu \nu} \square - \nabla_{\mu} \nabla_{\nu} \Bigl) G_k^{-1}. 
\end{align}
In what follow, we will consider a background without any matter source (i.e. $T_{\mu \nu}^{M} = 0$) in order to focus our attention on any effects, if it is present, given by the scale--dependent scenario.
To complete the set of equations, we can vary the effective action with respect to the scale-field $k(x)$ to produce
\begin{align}
\label{eomk}
\left[R \frac{\partial}{\partial k} \left(\frac{1}{G_k}\right)-
2 \frac{\partial}{\partial k}\left(\frac{\Lambda_k}{G_k}\right)\right]\cdot \partial k
=0.
\end{align}
While the above equation close the system, the implementation of this is a difficult task. Besides, the previous equations are complemented
by the relations corresponding to global symmetries of the system. Precisely, 
for the case of coordinate transformations we have
\begin{align}
\nabla^\mu G_{\mu \nu}=0.
\end{align}
In the next section we will briefly discuss a new black hole solution in the context of scale--dependent couplings inspired by quantum gravity recently reported in \cite{Rincon:2018lyd}.

\section{Black hole Solutions} \label{Model}
Adopting circular symmetry, and only radial dependence on the metric functions, we have the line element defined in terms of the usual Schwarzschild coordinates ($ct, r, \phi$) according to
\begin{align}\label{lineelans}
\mathrm{d}s^2 &= -f(r)\, \mathrm{d}(c t)^2 + \frac{\mathrm{d}r^2}{f(r)} + r^2 \Bigl[N(r)\mathrm{d}(c t) + \mathrm{d}\phi\Bigl]^2.
\end{align}
Solving the corresponding effective Einstein field equations, combined with the line element \eqref{lineelans}, we then have the scale--dependent functions given by
\begin{align} 
G(r) = &\frac{G_0}{1 + \epsilon r},
\label{G}
\\
N(r) = &- \frac{4 G_0 J_0}{c^3 r^2} Y(r),
\label{N}
\\
f(r) = &-8 \frac{G_0 M_0}{c^2} Y(r) + \frac{r^2}{\ell_0^2} + \frac{16 G_0^2 J_0^2}{c^6 r^2}Y(r)^2,
\label{f}
\end{align}
\begin{eqnarray}
\nonumber\Lambda(r)&&=-\frac{1}{\ell_0^2}\left(\frac{1+3\,\epsilon\, r}{1+\epsilon r}\right)-\frac{N^2(r)}{4}\,\left(\frac{\textrm{d}\ln Y(r)}{\textrm{d}\ln r}\right)^2+\\\nonumber
&&+\frac{8M_0 G(r) Y(r)}{c^2 r^2}\left[r \epsilon+\frac{1+2 r \epsilon}{2} \left(\frac{\textrm{d}\ln Y(r)}{\textrm{d}\ln r}\right)\right]+\\
&&-N^2(r)\frac{r \epsilon}{1+r\epsilon}\,\left(\frac{\textrm{d}\ln Y(r)}{\textrm{d}\ln r}\right).\label{lambda}
\end{eqnarray}

where
\begin{align}\label{Yder}
Y(r) \equiv 1 - 2 r \epsilon + 2 (r \epsilon)^2 \ln \bigg(1 + \frac{1}{r \epsilon}\bigg).
\end{align}

This solution is parametrized by five constants of integration, which are $\{ G_0, J_0, M_0, \Lambda_0, \epsilon \}$. 
Their physical meaning is directly linked to the classical solution as well as the non--rotating case. 
On one hand, the parameter $J_0 \rightarrow 0$ does not appear in the scale--dependent but non--rotating solution \cite{Koch:2016uso}. Taking this into account, we must impose that, for $J_0 \rightarrow 0$, the solutions \eqref{G},\eqref{N},\eqref{f} and \eqref{lambda} reduces to the solution reported in \cite{Koch:2016uso}
On the other hand, the rotating classical solution \cite{Banados:1992wn,Banados:1992gq},
should be obtained when the running parameter $\epsilon$ is taken to be zero, i.e. 
\begin{align}
\lim_{\epsilon \rightarrow 0} G(r) &= G_0,
\\
\lim_{\epsilon \rightarrow 0} N(r) &= N_0(r) \equiv -\frac{4 G_0 J_0 }{c^3 r^2},
\\
\lim_{\epsilon \rightarrow 0} f(r) &= f_0(r) \equiv -8\frac{ M_0 G_0}{c^2} + \frac{r^2}{\ell_0^2} + \frac{16 G_0^2 J_0^2}{c^6 r^2},
\\
\lim_{\epsilon \rightarrow 0} \Lambda(r) &= \Lambda_0.
\end{align}
Since corrections due to quantum scale dependence should be small, we expand the functions to first order in the $\epsilon$ parameter which implies that the metric functions are given by
\begin{align}\label{g}
G(r)& \approx G_0 \,(1 - \epsilon r),
\\
\label{n}
N(r) & \approx N_0(r) \,(1 - 2 \epsilon r),\\
\label{fex}
f(r) & \approx f_0(r)(1 - 4 \epsilon r) - 4 \epsilon r \Bigg[\frac{4 G_0 M_0}{c^2} - \bigg(\frac{r}{\ell_0}\bigg)^2 \Bigg],
\\
\label{lam}
\Lambda(r)& \approx \Lambda_0 (1+2r\epsilon).
\end{align}

\section{Sagnac Effect}
\label{Sagnac_effect}
\begin{figure}[h!]
	\begin{center}
		\includegraphics[width=65mm]{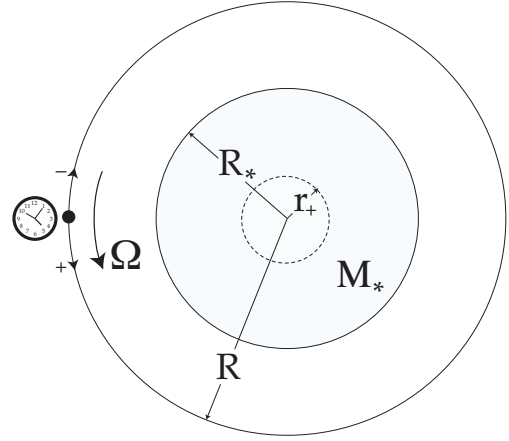}
	\end{center}
	\caption{Schematic of the Sagnac effect as measured by an observer with a clock rotating with constant angular velocity $\Omega$. The proper time difference between the emission and absorption of the co--propagating (+) and the counter--propagating (-) beam is considered.}
	\label{sbtz}
\end{figure}

In order to investigate the Sagnac effect, we consider a 2+1 circularly symmetric distribution of radii $r=R_*>r_+$ and classical mass $M_0=M_*$ embedded in an exterior scale-dependent BTZ background, as is shown in Fig. \ref{sbtz}. Rotating and static  fluid distributions that matches with the classical BTZ can be found, for example, in \cite{gurses,Cruz:1994ar,garcia03,Cruz:2004tz,Garcia:2004jz,cataldo04,Cataldo:2006yk,wuxu}. 
Assuming that the coordinates associated with the distribution are ($t', r', \phi'$),
we perform a change the coordinate to a another frame fixed in an interferometer that rotates at $r'=R>R_*$. 

Thus,
the transformation to the frame of the rotating platform 
\begin{equation}
\label{plat}c t=c t',\quad \phi=\phi'-\Omega t',
\end{equation}where $\Omega$ is the constant angular velocity of the physical frame, yields to the following metric coefficients
\begin{align}\nonumber
g_{tt}&= -f(R)+R^2\left(N(R)+\frac{\Omega}{c}\right)^2,
\\ \label{lineelans2}
g_{\phi \phi}&=R^2\\
\label{lineelans3}
g_{\phi t}&=g_{t \phi}=R^2\left(N(R)+\frac{\Omega}{c} \right).
\end{align}
Therefore, the non-zero components of the unit vector $\gamma^{\alpha}$ along the trajectory $r=R$ are given by 
\begin{eqnarray}
\label{gammafunc1} \gamma^t&=&\frac{1}{\sqrt{-g_{tt}}}=\gamma_J,\\
\label{gammafunc2} \gamma_t&=&-\sqrt{-g_{tt}}=-\gamma_J^{-1},\\
\label{gammafunc3} \gamma_{\phi}&=&g_{\phi t}\,\gamma^t=R^2\left(N(R)+\frac{\Omega}{c} \right)\,\gamma_J,
\end{eqnarray}
where
\begin{equation}
\label{gamj}\gamma_J=\left[f(R)-R^2\left(N(R)+\frac{\Omega}{c} \right)^2\right]^{-1/2},
\end{equation}
and thus, the gravito-magnetic vector potential becomes
\begin{equation}
\label{gmpot1} \Tilde{A}_{\phi}^G=c^2 \frac{\gamma_{\phi}}{\gamma_t}=-c^2 R^2 \left(N(R)+\frac{\Omega}{c} \right)\,\gamma_J^2.
\end{equation}

As was shown in \cite{Tartaglia:1998rh,Rizzi:2003uc,Rizzi:2003bf,Ruggiero:2005nd,Sultana:2013apa},
it is possible to express the phase shift $\Delta\Phi$ and time delay $\Delta \tau$ between light beams detected by a co-moving observer on the interferometer in terms of the gravito-magnetic vector potential by means of the expressions
\begin{equation}
\label{delph}\Delta \Phi
=\frac{2\, E_{\gamma}\, \gamma_t}{\hbar\,c^3}\int_{\zeta(S)}\vec{A}^G \cdot {\rm d}\vec{r},
\end{equation}and
\begin{equation}
\label{delt}\Delta \tau
=\frac{2 \gamma_t}{c^3}\int_{\zeta(S)}\vec{A}^G \cdot {\rm d}\vec{r},
\end{equation}where $E_{\gamma}$ is the relative energy of the photon as measured in the interferometer.

Then, the phase shift turns out to be
\begin{equation}
\label{shift1}\Delta \Phi=\frac{\tau_0\, E_{\gamma}}{\hbar} \frac{\Omega-\Omega_0}{\sqrt{\Omega_{\infty}^2-(\Omega-\Omega_{0})^2}},
\end{equation} and the (dimensionless) time delay becomes
\begin{equation}
\label{td1}\Delta \tilde{\tau}\equiv\frac{\Delta\tau}{\tau_0}= \frac{\Omega-\Omega_0}{\sqrt{\Omega_{\infty}^2-(\Omega-\Omega_{0})^2}},
\end{equation}where $\tau_0\equiv \frac{8\pi R}{c}$, $Y_R\equiv Y(R)$, and
\begin{align}
\label{ominf}\Omega_{\infty}&=\sqrt{\Omega_{\ell}^2+\Omega_{0}^2-\frac{8\,G_0 M_{*}}{R^2}Y_R},\\
\label{omla}\Omega_{\ell}&=\frac{c}{\ell},\\
\label{om0}\Omega_0&=4\frac{ G_0\,J_0}{c^2\,R^2} Y_R\equiv 4 \,\Omega_J Y_R.
\end{align}
Note that $\Omega_J$ can be interpreted as the precession velocity for the Lense--Thirring effect in the 2+1 gravity \cite{Lense:1918zz,Mashhoon93}:
\begin{equation}
\label{lteff}\omega_{LT}=-\Omega_J=-\frac{ G_0\,J_0}{c^2\,R^2},
\end{equation}which allow us to identify $\Omega_{LT}=-\Omega_0$ as the {\it precession velocity for the Lense--Thirring effect} for the scale--dependent rotating BTZ black hole. Also, by inspection of Eq. (\ref{td1}), we see that if $\Omega=0$ a time difference appears which is due to the rotation of the source. Obviously, if $\Omega=\Omega_0$ there is not Sagnac effect and an observer measure no time delay.

\begin{figure*}[ht]
	\centering
	\includegraphics[width=0.32\textwidth]{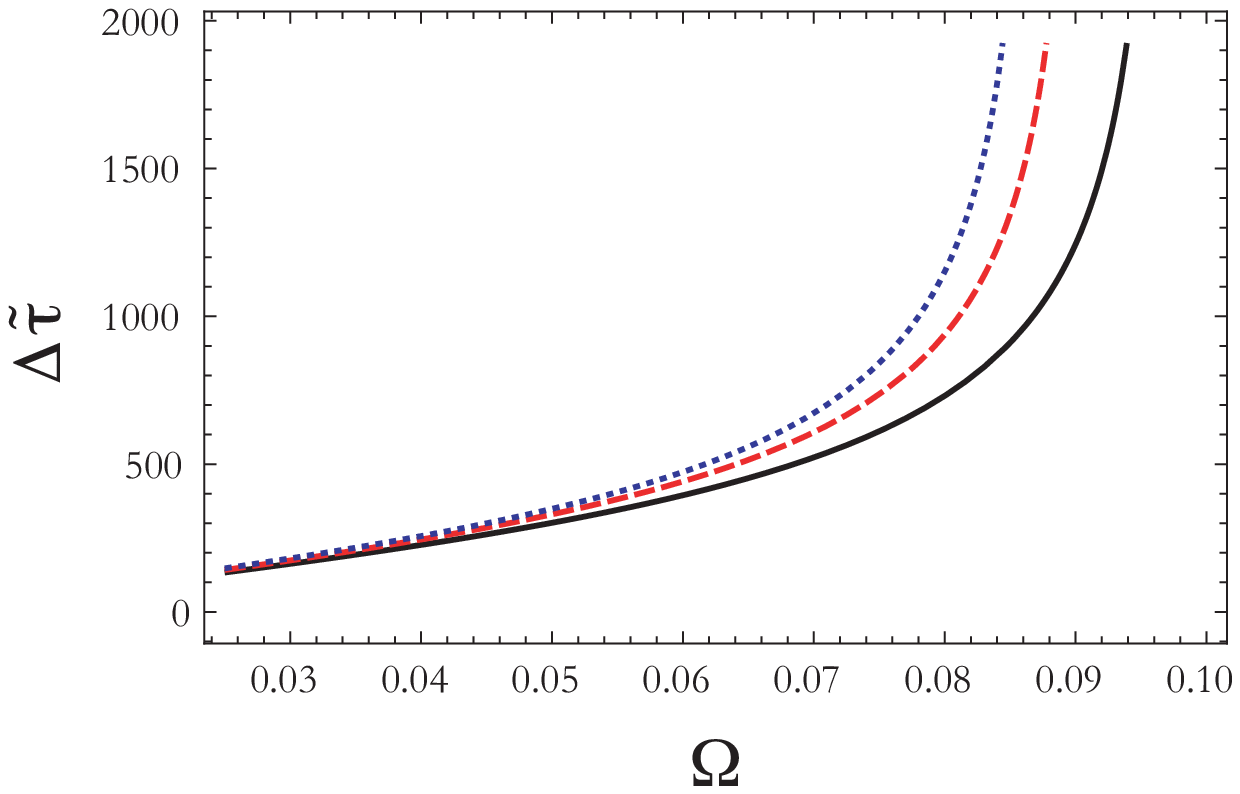}   \
	\includegraphics[width=0.32\textwidth]{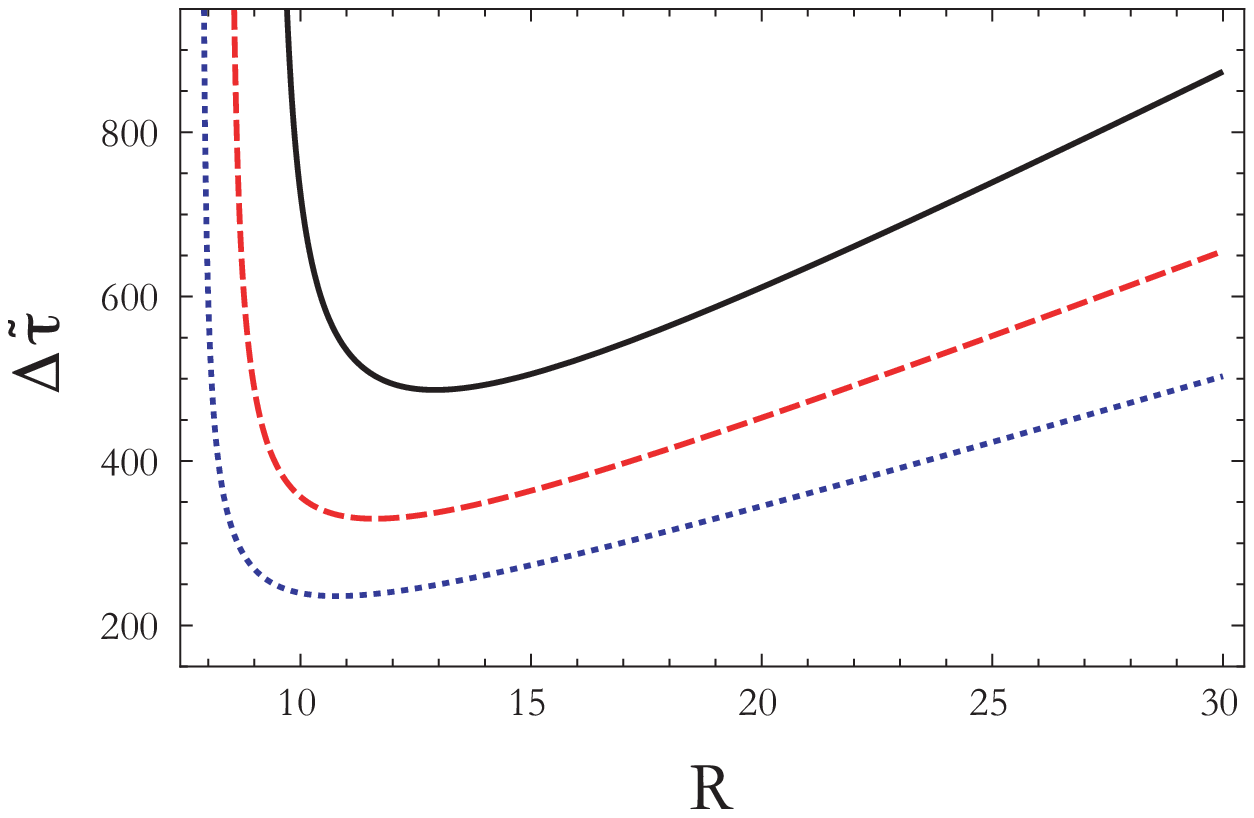}   \
	\includegraphics[width=0.32\textwidth]{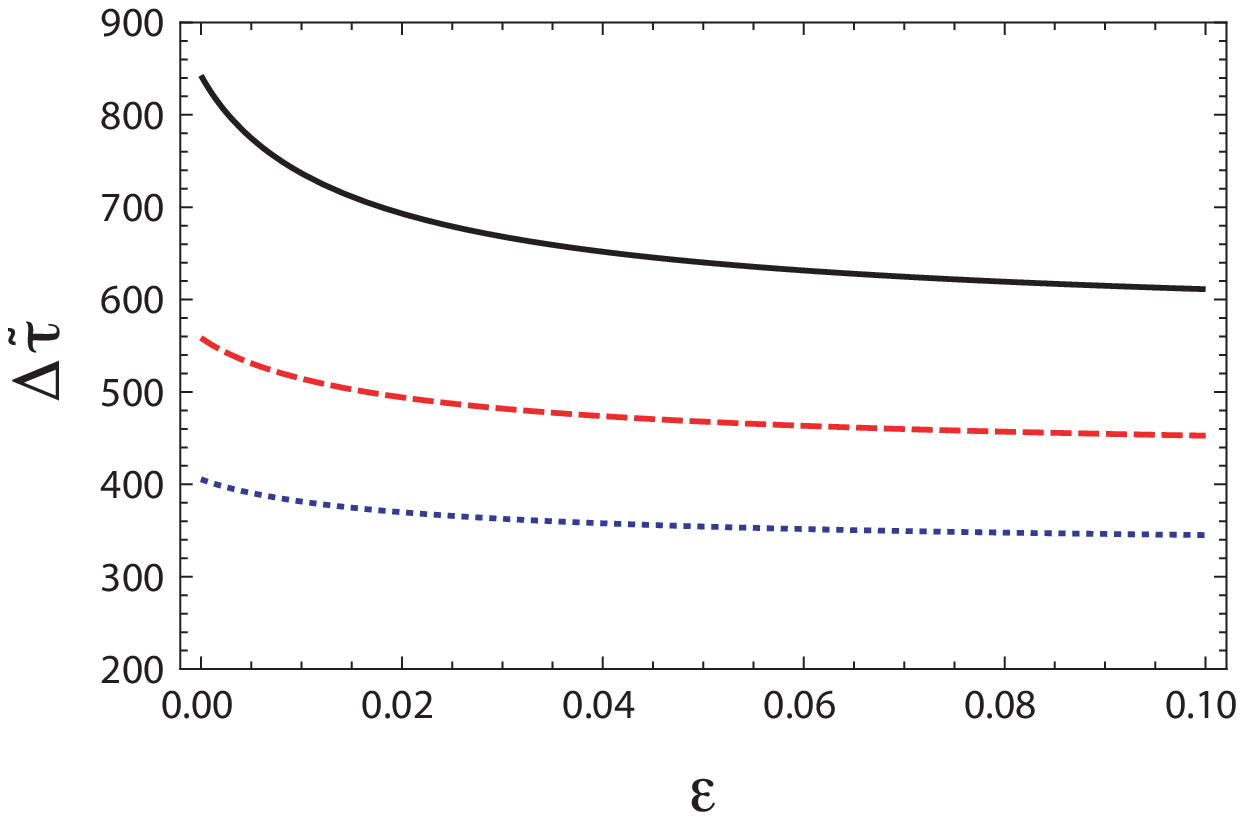}   \
	\caption{LEFT: Plot for the dimensionless time delay $\Delta \tilde{\tau}$ as a function of the angular velocity $\Omega$ for different values of the running parameter: black solid line $\epsilon=10^{-1}$, red dashed line  $\epsilon=10^{-2}$ and blue dotted line $\epsilon=10^{-5}$ (in arbitrary reciprocal length units) and fixed radii of the interferometer $R=20$ (in arbitrary length units).
		MIDDLE: Plot for the dimensionless time delay $\Delta \tilde{\tau}$ as a function of the angular velocity $\Omega$ for different values of the running parameter: black solid line $\epsilon=10^{-1}$, red dashed line  $\epsilon=10^{-2}$ and blue dotted line $\epsilon=10^{-5}$ (in arbitrary reciprocal length units) and fixed radii of the interferometer $R=20$ (in arbitrary length units). RIGHT: This plot shows the dimensionless time delay $\Delta \tilde{\tau}$ as a function of the running parameter $\epsilon$ for fixed orbital radii $R=20$ (in arbitrary length units) and different values of the angular velocity: black solid line $\Omega=7.5\times 10^{-2}$, red dashed line $\Omega=6.5\times 10^{-2}$ and blue dotted line $\Omega=5.5\times 10^{-2}$ (in arbitrary reciprocal time units).}
	\label{taugen}
\end{figure*}

In the left panel of Fig. (\ref{taugen}) we have plotted the dimensionless time delay $\Delta\tilde{ \tau}\equiv \Delta \tau /\tau_0$ as a function of the angular velocity of the interferometer $\Omega$ for different values of the running parameter $\epsilon$, showing that the increase in $\epsilon$ causes a decrease in the maximum angular velocity that the device can reach to obtain the Sagnac effect. 
In the middle panel of Fig. \ref{taugen} we have plotted the dimensionless time delay $\Delta\tilde{\tau}$ as a function of the orbital radii of the interferometer $R$, for fixed value of the running parameter $\epsilon=10^{-1}$ (in arbitrary reciprocal length units) and different values of the angular velocity: $\Omega=7.5\times 10^{-2}$, $\Omega=6.5\times 10^{-2}$ and $\Omega=5.5\times 10^{-2}$ (in arbitrary reciprocal time units). This plot shows that there is minimum for $\tilde{\tau}$, which depends strongly on  the running parameter such that the position of this minimum grown directly with $\epsilon$.
In the right panel of Fig. \ref{taugen} we shows the dimensionless time delay $\Delta \tilde{\tau}$ as a function of the running parameter $\epsilon$ for fixed orbital radii $R=20$ (in arbitrary length units) and different values of the angular velocity: $\Omega=7.5\times 10^{-2}$, $\Omega=6.5\times 10^{-2}$ and  $\Omega=5.5\times 10^{-2}$ (in arbitrary reciprocal time units).

\begin{figure}[h!]
	\begin{center}
		\includegraphics[width=80mm]{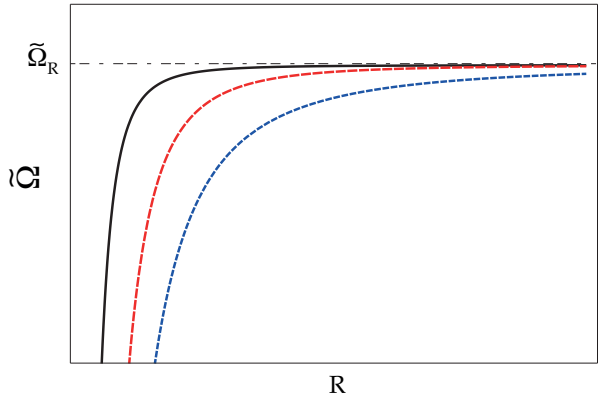}
	\end{center}
	\caption{Plot for the dimensionless angular velocity $\widetilde{\Omega}$ as a function of the orbital radii of the interferometer, $R$, for different values of the running parameter: $\epsilon=10^{-1}$, $\epsilon=10^{-2}$ and $\epsilon=10^{-5}$, in arbitrary reciprocal length units.}
	\label{sg1}
\end{figure}

On the other hand, for the non-rotating case, $\Omega_J=0$, the dimensionless time delay can be written as
\begin{equation}
\label{senr}\Delta \tilde{\tau}= \frac{\Tilde{\Omega}}{\sqrt{1-\frac{Y_R}{Y_+}\,\left(\frac{r_+}{R}\right)^2-\widetilde{\Omega}^2}},
\end{equation}where $\widetilde{\Omega}\equiv \Omega/\Omega_{\ell}$ is the dimensionless angular velocity,  $Y_+\equiv Y(r_+)$, and $r_+$ is the horizon event of the non-rotating scale--dependent BTZ black hole obtained from Eq. (\ref{f}) for $J_0=0$.  As is expect in this case, $\Omega=0$ implies that no Sagnac effect arise. Also, note that there is an upper limit for the (dimensionless) angular velocity $\widetilde{\Omega}_R$ given by
\begin{equation}
\label{uppl}\widetilde{\Omega} <\widetilde{\Omega}_R\equiv \sqrt{1-\frac{Y_R}{Y_+}\,\left(\frac{r_+}{R}\right)^2}.
\end{equation}In Fig. \ref{sg1} we have plotted $\widetilde{\Omega}$ as a function of $R$ for three values of the running parameter $\epsilon=10^{-1}$, $\epsilon=10^{-2}$ and $\epsilon=10^{-5}$ in arbitrary reciprocal length units, and shows that the increase in $\epsilon$ causes the maximum angular velocity $\widetilde{\Omega}_R$ to be reached faster.

An interesting remark for the last result is that in the classic non-rotating BTZ black hole the last relation reduces to
\begin{equation}
\label{upplnr}\widetilde{\Omega} <\widetilde{\Omega}_R\equiv \sqrt{1-\left(\frac{r_+}{R}\right)^2}.
\end{equation}

Recently, a similar result was obtained by Villanueva et al. for a 3+1 static toroidal topological black hole in conformal Weyl gravity \cite{Villanueva:2018kem}, where the term $(r_+/R)$ has a power index 3 instead the index 2 present in Eq. (\ref{uppl}). 
This result looks dependent on the spatial dimension of the manifold, but that assertion deserves a more depth study.

Finally, note that for the classical rotating BTZ black hole, the time delay obtained by Raychaudhury \cite{RAYCHAUDHURI:2014sma} contain a sign mistake in his Eq. (18), since the quantity inside of the square root becomes negative for $r>r_+$. Thus, in the limit $\epsilon\rightarrow 0$ ($Y\rightarrow 1$), Eq. (\ref{td1}) reduces to the correct expression for the Sagnac effect for the classical rotating BTZ black hole.

\section{Conclusions} \label{Conclusions}

In this article, we have studied the Sagnac effect in light of the scale--dependent scenario in a rotating BTZ black hole background. As our formalism incorporate quantum corrections, our solution generalizes the standard case previously reported in \cite{RAYCHAUDHURI:2014sma}. It is remarkable that we should recover the classical Sagnac effect when we turn $\epsilon$ off, however it is not the case. We thus conclude that the Raychaudhuri solution (and not our solution) has a minimal error. More precisely, the discrepancy appears when we analyze the global sign inside the square, resulting in a non-physical solution for a Sagnac effect because the validity of his expression is for $r<r_+$ (Eq. (18) in his paper).
An interesting features is present in our solution, namely, in the scale--dependent scenario the time delay decreases when the running parameter increases, effect which can gives some insights about the presence of quantum effects. Thus, it provides a new way to check if the corresponding quantum corrections modifies the underlying theory.
Also, an important remark about the non-rotating case, more precisely  inside the square root of Eq. (\ref{upplnr}), is that upper limit of the angular velocity depends on the term $(r_+/R)^2$ instead of  $(r_+/R)^3$, which was found by Villanueva et al. \cite{Villanueva:2018kem} in the context of the topological toroidal black hole in the 3+1 conformal Weyl gravity. This fact looks depends on the number of spatial coordinates, but that assertion deserves a more depth investigation.

\begin{acknowledgements}
	The authors acknowledge useful conversations with Prof. Dr. Michel Cur\'e and Prof. Dr. Marco Olivares.
	The work of A.R. was supported by the CONICYT--PCHA / Doctorado Nacional / 2015--21151658 and the work of J.V. was partially supported by the Centro de Astrof\'isica de
	Valpara\'iso (CAV).	
\end{acknowledgements}

\bibliographystyle{spphys} 
\bibliography{Biblio_v1.bib}

\end{document}